\documentclass{svjour2}                    % onecolumn
\smartqed  % flush right qed marks, e.g. at end of proof
\usepackage{graphicx}
\newcommand{\bu}{{\bf u}}  
\newcommand{\bx}{{\bf x}}  
\newcommand{\br}{{\bf r}}  

\newcommand{\ee}{\end{equation}}
\newcommand{\be}{\begin{equation}}

\begin{document}

\title{Fully developed turbulence and the multifractal conjecture}

%\subtitle{Do you have a subtitle?\\ If so, write it here}

%\titlerunning{Short form of title}        % if too long for running head

\author{Roberto Benzi        \and
        Luca Biferale %etc.
}

%\authorrunning{Short form of author list} % if too long for running head

\institute{Luca Biferale\at Dept. Physics and INFN, Universtiy of Tor Vergata, via della Ricerca Scientifica 1 
               \\
              Tel.: +390672594595\\
              Fax: +39062023507\\
              \email{biferale@roma2.infn.it}           %  \\
%             \emph{Present address:} of F. Author  %  if needed
          % \and
          % S. Author \at
          %    second address
}

\date{Received: date / Accepted: date}
% The correct dates will be entered by the editor

\maketitle

\begin{abstract}

 We review the {\it Parisi-Frisch } \cite{ParisiFrisch} MultiFractal formalism for
  Navier--Stokes turbulence with particular emphasis on the issue of
  statistical fluctuations of the dissipative scale.  We do it for
  both Eulerian and Lagrangian Turbulence. We also show new results
  concerning the application of the formalism to the case of Shell
  Models for turbulence. The latter case will allow us to discuss the
  issue of Reynolds number dependence and the role played by vorticity
  and vortex filaments in real turbulent flows.
\keywords{Eulerian and Lagrangian Turbulence\and Multifractals}
% \PACS{PACS code1 \and PACS code2 \and more}
% \subclass{MSC code1 \and MSC code2 \and more}
\end{abstract}

\section{Introduction}
Turbulent flows are characterized by strong fluctuations of energy
transfer from small to large scales \cite{Fr.95}. 
The best known theoretical
approach to describe the statistical properties of turbulent flows is
due to Kolmogorov in 1941 (K41).  The K41 theory assumes that
turbulent fluctuations can be considered homogeneous and isotropic at
very small scales. Moreover, Kolmogorov assumed that the average (in
time) rate of energy dissipation $\epsilon$ is Reynolds independent
for large enough Reynolds number, $Re \equiv U_0 L_0/\nu$, where $U_0$ is some
characteristic velocity at large scale $L_0$ and $\nu$ is the kinematic
viscosity of the flow. For the sake of simplicity we will set $U_0=L_0=1$ from now on. Thus $Re = \nu^{-1}$. The equations of motions governing
hydrodynamical incompressible turbulence are the Navier--Stokes
equations, with density set to $\rho=1$:
\begin{equation}
\label{eq:ns}
\partial_t \bu + (\bu {\bf \, \partial_r}) \bu = {\bf \partial_r} p + \nu \Delta \bu,
\end{equation}
where $u_i$ is a three-dimensional field satisfying $\partial_i u_i =
0$.  Note that in terms of $u_i$ the average rate of energy
dissipation is given by $\epsilon \equiv \sum_{ij}\nu (\partial_i u_j)^2$.
Thus, $\epsilon$ becomes independent of $Re$ if at
large Reynolds number $\partial_i u_j \sim Re^{1/2}$.

The statistical properties of the vector field $u_i$ can be studied by
using the $n$-order tensor, given by the  correlation functions $\langle
u_{i_1}(\bx_{a_1}) ...u_{i_n}(\bx_{a_n}) \rangle$, where the
index $i_1,..i_n$ stands for the spatial directions and
$\bx_{a_i}$ are $n$-points in the physical space, and the symbol
$\langle ...\rangle$ means ensemble average.  If we assume, following
Kolmogorov, that turbulence becomes isotropic and homogeneous at
scales small enough, then we can rewrite the correlation functions in
terms of only longitudinal velocity increments, $\delta_r u = \delta
\bu(\br) \cdot {\hat \br} $ (where $\delta \bu(\br) =
\bu(\bx+\br)-\bu(\bx)$) and transverse increments, $ \delta_r w =
\delta {\bu}({\br}_T) $, (where ${\br}_T \cdot {\bu} = 0$).  For a
discussion on the complexities arising when also anisotropic
fluctuations are relevant see \cite{bp}.

One of the main predictions of the Kolmogorov theory is contained in his
famous $4/5$-equation for homogeneous and   isotropic turbulence:
\begin{equation}
\label{K45}
\langle (\delta_r u)^3 \rangle = -\frac{4}{5} \epsilon r + 6 \nu \frac{d}{dr} \langle (\delta_r u)^2 \rangle.
\end{equation}
For large $Re$ (i.e. small $\nu$), and at fixed distance $r$, the $4/5$ equation tells us two important information: the statistical properties of $\delta_r u$ have a non zero skewness, and they have a  scaling  behavior as a 
functions of $r$. Note that (\ref{K45}), taken in the limit of
vanishing viscosity, is dimensionally consistent with having $\delta_r
u  \sim r^{1/3}$, which 
implies for the $p$-th order  longitudinal Structure Functions,  $S^{(p)} (r) \equiv \langle (\delta_r u)^p \rangle \sim r^{p/3}$.

It turns out experimentally that the K41 theory is only partially correct,
i.e. equation (\ref{K45}) is very well verified when the reference
scale, $r$, is much smaller than the integral scale, $L_0$, implying a
recovery of homogeneity and isotropy for small
enough scales. However, the structure function data are not in agreement with the Kolmogorov
theory for all orders $p$.  For instance, K41 theory predicts that
the hyper-Flatness, $S^{(p)}(r)/(S^{(2)}(r))^2$, should be constant as
a function of $r$, while it is not. Both experiments and numerics show
that it grows  when going from large to small scales. This
effect is usually referred to as ``intermittency'' and it is clearly
missing in the K41 scenario (for a collection of experimental and
numerical data on both Eulerian and Lagrangian data, see
\cite{arneodoetal96,arneodoetal08}).

What is the physical explanation of intermittency? Is intermittency a
finite Reynolds effect, i.e. does it disappear asymptotically for
large $Re$?  Are scaling properties of the turbulent field spoiled by
intermittency? These questions are the fundamental issues discussed
by the scientific community in the last 50 years.  Unfortunately,
there are no exact results about intermittency in Navier--Stokes
turbulence. So, we must rely both on experimental and numerical data,
or on phenomenological and dynamical modeling of turbulence. In
particular, a key problem is to control the effects of viscosity,
i.e. how to connect experimental and numerical data -- naturally limited
in Reynolds numbers -- to any theoretical or phenomenological
understanding proposed for the asymptotic infinite Reynolds number
case. In this paper we will deal with such an issue, focusing on the
MultiFractal (MF) phenomenological description of ``infinite Reynolds''
turbulence and on its extension to the case of finite Reynolds.

The paper is organized as follows. First, we review some of the
results so far obtained by employing the MultiFractal theory of
turbulence, introduced 25 years ago by Parisi and Frisch
\cite{ParisiFrisch,Fr.95} (see \cite{review.angelo} for a recent
review). Then, we discuss subtleties connected on how to introduce in
a self-consistent way the effects of viscosity in the MF description,
i.e. how to control finite Reynolds effects for both Eulerian and Lagrangian turbulence. Finally, we present new
data obtained on a class of dynamical systems for the turbulent energy
cascade, the so-called Shell Models \cite{Fr.95,PalVul,lb.arfm}. This will allow us to address and test the MF formalism at 
changing Reynolds numbers.  We conclude with some perspectives for
further work.

\section{Review of the multifractal approach of turbulence}
\label{sec:mf}
The multifractal approach to turbulence is based on the assumption
that the statistical properties of turbulent flows do exhibit scaling
properties even if there is intermittency.  Let us discuss scaling in
a rather more abstract way by using the symmetries underlying 
Euler and  Navier--Stokes equations.  The Euler equations have the
``standard'' symmetries of the Newton laws for inviscid fluid,
i.e. isotropy, parity and time reversal. Moreover, the Euler equations
display another important set of symmetries of global scale invariance,
i.e. invariance under the transformations \cite{frisch85}:
%\cite{frisch-les-houches}
\begin{eqnarray}
r \rightarrow \lambda r;\quad u \rightarrow \lambda^{h} u;\quad t \rightarrow \lambda^{1-h} t.
\end{eqnarray}
The value of $h$ is arbitrary and it is not fixed by any physical
and/or mathematical constraint. The Navier--Stokes equations have no
time reversal (due to viscosity) and satisfy a new scale invariance
property, namely:
\begin{eqnarray}
\label{scale}
 r \rightarrow \lambda r; \quad
 u \rightarrow \lambda^{h} u; \quad
 t \rightarrow \lambda^{1-h} t;\quad
 \nu \rightarrow \lambda^{1+h} \nu.
\end{eqnarray}
Note that the new scaling property implies $ \epsilon \rightarrow
\lambda^{3h-1} \epsilon.$ We can now rephrase the K41 theory in the
following way: using (\ref{K45}) the energy dissipation rate should
be constant and equal to $S^{(3)}(r)/r$ which implies $h=1/3$.  In
other words, the scaling transformation $r \rightarrow \lambda r$
($\lambda \ll 1$) is equivalent to a change in the viscous effect and,
therefore, to a change of $\epsilon$. If we assume
that $\epsilon $ is homogeneous in space and time, it follows that $h$
cannot be arbitrary and should be fixed to $h=1/3$.  Let us note that the scale
invariance for the Navier--Stokes equations is now restricted to a
particular value of $h$, i.e. there is less ``scale'' symmetry in the
system as naively predicted by the equation of motions. However, one
can take a different point of view, namely one can assume that $h$ is
still arbitrary, i.e.  the scale symmetry holds, although $\epsilon$
is constant only on average. The problem is to understand what is meant
by ``average''. This is the conceptual step performed in the
Parisi-Frisch paper \cite{ParisiFrisch}. The idea is to assume that in the limit
of infinite Reynolds numbers, and for any fixed scale, $r$,
the scale invariance holds with some probability $P_r(h)$. It is then
assumed that $\delta_r u$ and $P_h(r)$ are scaling function of $r$, i.e.
\begin{equation}
\label{eq:pf}
\delta_r u \sim r^h; \qquad P_h(r) \sim r^{F(h)}.
\end{equation}
Then, all the correlation functions should be computed by averaging
over the probability $P_h(r)$:
\begin{equation}
\label{zetap}
S^{(p)}(r) \propto  \int dh \  r^{F(h) + ph}.
\end{equation}
The breaking of time reversal still leads to $\epsilon \sim const$
which implies the constrain $S^{(3)}(r)/r \sim const$. Using the
definition: $$\zeta(p) \equiv \min_h[F(h) + ph],$$ we can perform the integral by
the saddle point method and obtain $S^{(p)}(r) \sim r^{\zeta(p)}$ with the constraint
$\zeta(3) = 1$.

Historically, $F(h)$ has been written as $3-D(h)$ by assuming that
$D(h)$ is the fractal dimension of the set at scale $r$ where
$\delta_r u \sim r^h$ and, for this reason, the above picture has been
named the ``multifractal'' approach to turbulence. We note that there is no
reason to introduce any fractal dimension, i.e. there is no {\it a
  priori} geometrical interpretation in defining $P_h(r)$.

The crucial physical point, which is discussed in this paper, is the
role of the viscosity in the multifractal approach. The scaling
invariance of the Navier--Stokes equations tells us that by a scale
transformation $ r \rightarrow \lambda r $ nothing changes if the
viscosity becomes smaller by a factor $\lambda^{1+h}$. In the real
world, the viscosity is fixed and therefore it increases as
$\lambda^{-(1+h)}$.  Eventually, the viscosity becomes so large as to
kill any turbulent fluctuations. We can assume that viscosity is
relevant if the turbulent characteristic time scale $r/\delta v(r) $
is of the same order of the viscous time scale $r^2/\nu$, which
defines a fluctuating, i.e. $h$-dependent, dissipation scale $\eta(h)$:
\begin{equation}
\label{fv}
\frac{\eta \,\delta_\eta u }{\nu} \sim 1 \ \rightarrow \eta(h) \sim \nu^{1/(1+h)}.
\end{equation}
The idea of a fluctuating dissipation scaling has been introduced by
Paladin and Vulpiani \cite{PV.87} and it should be considered as a
consequence of the scale invariance (\ref{scale}).  Let us discuss in
more details this point.

A possible interpretation of the scale invariance would suggest that the
velocity difference $\delta_r u$ -- and this has implications for the $n$-point
correlation functions -- depends on $h$ in the following manner: 
 \be \delta_r u \sim g(r/\eta_{k41})
f(r/\eta_{k41}))^h,\ee
where $\eta_{k41}$ is the Kolmogorov --not
fluctuating-- dissipative scale estimated as $\eta_{k41} \equiv
(\nu^3/\epsilon)^{1/4}$. The appropriate asymptotic behaviors for the functions $g(x)$ and $f(x)$ 
must be: $g(x) \sim const$ and $f(x) \sim x$ for $x \gg 1$, while $g(x) \sim x$ and 
$f(x) \sim const$ for $x \ll 1$.
%The function $f(x)$ must go to a constant, $f(x)
%\sim const$, for $x \gg 1$, if we want to recover the infinite
%Reynolds behaviour, while for $r \ll \eta_{k41}$ we should expect
%$\delta_r u \sim r$, and therefore $g(x) \sim const $  and $f(x) for $x \ll
%1$. 
Thus $f(x)$ and $g(x)$ can be interpreted as  cutoff functions on the
inertial-range scaling behavior introduced by the dissipation.  Let us
remark that we can also expect a cutoff function entering into the
definition of the probability distribution $P_h(r)$, i.e. we should
expect that \be
\label{ESS}
P_h(r) \sim [ f(r/\eta_{k41})]^{F(h)}.  \ee 
It is interesting to remark that  in the range
of scales where \be \label{eq:fg} g(x) \sim {\rm const.} \ee we have, using
(\ref{ESS}) that $S^{(p)} (r) \sim [r\,
f(r/\eta)]^{\zeta(p)}$. Recalling that $\zeta(3)=1$, the previous
expression can be rewritten as $S^{(p)}(r) \sim (S^{(3)}(r))^{\zeta(p)}$.
It turn out that the range of scales where the equality (\ref{eq:fg})
holds is much more extended then the usual inertial range, i.e. from
experimental and numerical data one can safely deduce that the
relation (\ref{eq:fg}) is valid down to scales $ x \simeq 10$; i.e. where
viscosity is already important. As a result, both function $f(x)$ and
$g(x)$ are not constant anymore at those scales, but the viscous
corrections on both functions are almost the same.  This empirical
fact is known as ESS (Extended Self Similarity), it has been
extensively used in literature \cite{ess1,ess2} to estimate the
scaling exponents $\zeta(p)$ with good accuracy. Eq. (\ref{eq:fg}) cannot hold also for very small scales, $ x \ll 1$,
where viscosity starts to have a dominant role. In order to take into
account also of these effects, we need to consider relation (\ref{fv})
and its consequences.  The generalization can be done  following \cite{meneveau}
\begin{eqnarray}
\label{dvh}
\delta_r u \sim g(r/\eta(h)) (f_h(r/\eta(h))^h; \qquad P_h(r) \sim [ f(r/\eta(h))]^{F(h)}.
\end{eqnarray}
%Notice that the function $f_h(x)$ should depend {\it both} on $h$ and on $r/r_d(h)$, since the following limits must hold:
%\begin{eqnarray}
%\label{larger}
%&&f_h(r/r_d(h)) \rightarrow 1 \ \ \ for \ \ \ r/r_d(h) \rightarrow \infty \\
%&&f_h(r/r_d(h)) \rightarrow [\frac{r}{r_d(h)}]^{1-h} \ \ \ for \ \ \  r/r_d(h) \rightarrow 0 
%\label{smallr}
%\end{eqnarray}
In section(\ref{sec:bm}), we will discuss some explicit form of the
function $f(x)$ and $g(x)$.  Equations (\ref{dvh}) 
tells us how to introduce in a self-consistent way the effect of
finite viscosity, i.e. breaking of scale invariance) within the
scale invariant MF approach. We will refer to the whole picture as the
{\it multifractal conjecture}. More precisely, we must expect that for
all quantities that are invariant with respect to the Navier--Stokes
symmetries, the scaling properties can be obtained by using  the
same universal function $F(h)$, including the effect of finite
viscosity.  Thus, all the Reynolds number effects on the statistical
properties of turbulence should be predicted using the function
$F(h)$, the scaling invariance {\it and} equation (\ref{fv}).  From
this point of view, the multifractal conjecture should be able to
predict how different quantities (for instance moments of the velocity
gradients) depend on $Re$. The explicit functional form for the
interpolating functions $f(x)$ and $g(x)$ may be safely considered not
particularly relevant, as soon as the asymptotic scaling limits are preserved
(see also \cite{chevillardphysd,yakhot_old} for some preliminary
attempts to determine them using constraints from the Navier--Stokes
equations).

Let us make some remarks on the above picture, mostly in a 
historical perspective. From the theoretical point of view, the  major
challenge is to compute the function $F(h)$.  In some particular,
linear cases, this computation can be done \cite{gk}, although no one
knows how to perform such calculations for the Navier--Stokes
equations. There has been a lot of effort to clarify, using both
experiments and numerical simulations, whether the statistical
properties of turbulence exhibit scaling behavior. So far, there is a
wide consensus that this is the case. Knowing the exponents $\zeta(p)$
one can easily compute the function $F(h)$ by some reasonable
``fitting'' procedure. Thus, we can verify whether the multifractal
conjecture is satisfied or not, i.e. whether there exists or not some
quantity (invariant with respect the group of Navier--Stokes symmetry)
which cannot be predicted knowing the function $F(h)$.  
%Clearly, one
%can study something which is not invariant. For instance, one can
%study quantities like $\langle (\partial_x u_y)^2 \rangle$ (which is
%not invariant under rotation) as a function of $Re$ {\bf LUCA: questo
%  non lo capisco, che vuoi dire?}. 
For non-invariant quantities, the
multifractal conjecture tells us nothing. Also, there have been some
approaches in the past claiming that for large enough $Re$, the
statistical properties of turbulence are asymptotic to the K41
theory. Experiments and numerical simulations do not support
such claims \cite{arneodoetal96,arneodoetal08,Go.02}, so far.

\section{Predictions concerning the fluctuating dissipative scale}

\subsection{Eulerian Framework}
\begin{figure}
\centering
  \includegraphics[width=0.95\textwidth]{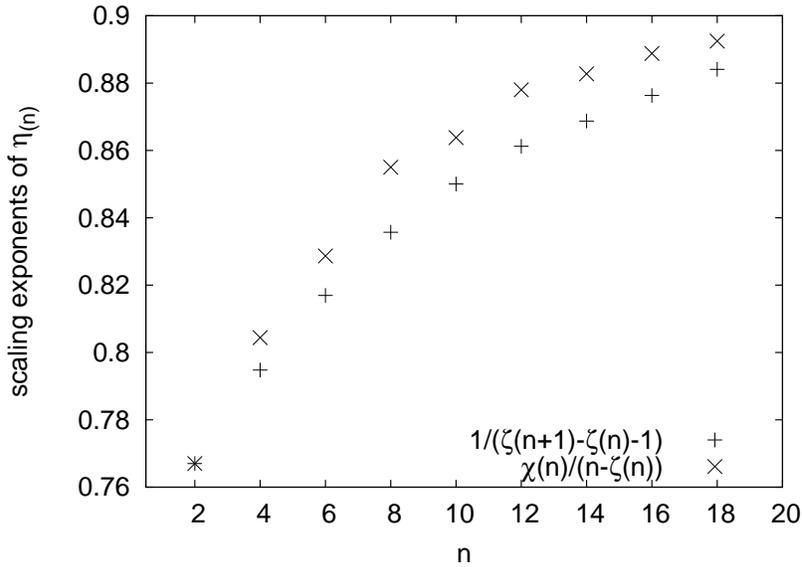}
\caption{Value of the exponents entering in the expressions of the
  moment dependent viscous cutoff either  from (\ref{eq:lb}),
  $\times$; or from (\ref{eq:vy}), $+$. Notice that the value for $n=2$ is the same, as it comes from the request $\zeta(3)=1$.}
\label{fig:0}
\end{figure}

The multifractal conjecture allows us to develop a number of different
predictions. Let us assume that we know the function $F(h)$ or, equivalently,
the scaling exponents $\zeta(p)$. It is then possible to predict the
scaling behavior of the following non trivial quantities. Some of
these predictions are based on the basic idea that the dissipation
scale of turbulence is a fluctuating quantity satisfying
eq. (\ref{fv}). It is then possible to compute the probability
distribution of the velocity gradients by using the estimate 
\cite{PV.87,Ne.90,Bif.91}:
\begin{equation}
\partial_r u \sim \frac{\delta u(\eta)}{\eta}.
\end{equation}
Since $\delta_{\eta(h)} u \sim (\eta(h))^h$ and $\eta(h) \sim
\nu^{1/(1+h)}$, we have: \be \langle(\partial_r u)^{q}\rangle \sim
\int dh \nu^{(q(h-1)+F(h)))/(1+h)} \quad \sim Re^{\chi(q)}, \ee which
predicts the scaling properties of the gradient as function of $Re
\sim 1/\nu$, and where the last expression is obtained by the saddle
point method when the Reynolds number $Re \sim \nu^{-1}$ tends to
infinity.
Here
$$\chi(q) \equiv \min_h[(2q\,(h-1)+F(h))/(1+h)].$$ Note that the prediction is
highly non trivial and it is connects $\chi(p)$ with the exponents
$\zeta(p)$ via a non-linear relation valid for any function $F(h)$,
(see \cite{Fr.95}):
\begin{equation}
\label{eq:grad}
\chi(q) = \frac{p-\zeta(p)}{2}; \qquad {\mbox with} \quad  q = \frac{\zeta(p)+p}{2}.
\end{equation}
The above relation implies  the existence of
dissipative anomaly, i.e. $\chi(2) = 1$,
 for any spectrum such that $\zeta(3)=1$:
\begin{equation}
\lim_{Re \rightarrow \infty} \epsilon \sim Re^{-1} \langle (\partial_r u )^2 \rangle \rightarrow
const. 
\end{equation}
Such non-trivial multifractal spectrum for velocity gradients may be
used to define a multifractal spectrum of dissipative scales, each-one
defining the different cross-over from the inertial range to the
dissipative range for different moments. To implement this, we may follow two
slightly different methods. First, let us define the $n$-th order
dissipative scale as the intersection between the extrapolation for
large scales of the differential smooth behavior of $S^{(n)}(r)$ with
the extrapolation for small scales of the inertial behavior:
\begin{equation}
 S^{(n)}(r) \sim r^n \langle (\partial_r u)^n \rangle \qquad {\mbox  if }
\quad 
r \ll \eta_{k41} \end{equation}
\begin{equation}
S^{(n)}(r) \sim r^{\zeta(n)} \qquad {\mbox if }
\quad 
r \gg \eta_{k41}.
\end{equation}
The intersection between the two power-law behavior can be considered
a good estimate for the typical dissipative scale, $\eta_{(n)}$,
of the  $n$-th order structure functions:
\begin{equation}
(\eta_{(n)})^n \langle (\partial_r u)^n \rangle = (\eta_{(n)})^{\zeta(n)}.
\end{equation}
Using expression (\ref{eq:grad}) for the scaling of
gradients, we get:
\be
\label{eq:lb}
\eta_{(n)} \sim Re^{-\chi(n)/(n-\zeta(n))}.
\ee
Notice that according to this definition we have for the dissipative
cut-off of the second order structure function:
\be
\eta_{(2)} \sim Re^{-\chi(2)/(2-\zeta(2))}.
\ee
Another, slightly different, way to define the moment-dependent viscous cut-off
is the one followed in refs. \cite{yakhot1,yakhot2}, where 
a  balancing inside the equations of
motion between  inertial and dissipative terms is used to obtain 
$\eta_{(n)}$. Following \cite{yakhot1,yakhot2}, 
it is easy to derive the dimensional balancing between the inertial
terms and the dissipative contribution in the evolution of the $n$-th
order generic structure functions,  from (\ref{eq:ns}):
\be
S^{(n+1)}(r)/r \sim \nu S^{(n)}(r)/r^2.
\ee
If the $n$-th order dissipative scale is now obtained by asking it to be
the scale where the previous matching holds, we easily obtain,
another expression for the cut-offs \cite{yakhot1}:
\be
\label{eq:vy}
\eta_{(n)}' \sim Re^{-1/(\zeta(n+1)-\zeta(n)+1)}.
\ee
Let us notice that the two expressions (\ref{eq:lb}) and (\ref{eq:vy})  are exactly equivalent for the
cut-off entering in the dissipative anomaly $\eta_{(2)} \equiv \eta_{(2)}' = Re^{-1/(2-\zeta(2))}$, for any choice of $F(h)$,  while they are slightly
different for  high-order moments. The discrepancies are however very
small, as can be directly checked by plugging some possible shape of
the $\zeta(p)$ exponents that fit the experimental data (for example
the log-Poisson expression derived in \cite{sl}, as shown in
fig.~\ref{fig:0}).
%%%%% FIGURE intro1 %%%%%%%%%%%%%%%
%%%%%%%%%%%%%%%%%%%%%%%%%%%%%%%%%%
Recently, ad-hoc advanced computations have been performed by changing
the numerical resolution, in order to test the existence of
such non-trivial fluctuations at  dissipative scales
\cite{shumacher}. The numerical data have also been
succesfully compared with the MF prediction (\ref{fv}) in \cite{lb1}. Note that it
is quite difficult to get good laboratory data for velocity
fluctuations at scales much smaller than $\eta_{k41}$ due to the
intrusive nature of the experimental probes. Nevertheless, in
recent years very interesting and promising experimental techniques
have been developed to track particles in turbulent flows, accessing
temporal fluctuations instead of spatial fluctuations
\cite{LPVCAB.01,MMMP.01,ott_mann,tsinober,toschibod,zellmann}.  The quality of
predictions based on the multifractal conjecture is even more striking
by performing the statistical analysis of the velocity field in the
Lagrangian framework.
\subsection{Lagrangian Framework} 
\label{sec:bm}
Given the importance of particle dynamics in turbulent flows, numerous
numerical and experimental studies have flourished in the last few
years \cite{MMMP.01,berg,X06,mueller,BBCLT.05,pkyeung,chicago} (see also \cite{toschibod} for a recent review). 
Neutrally buoyant particles, advected by a turbulent velocity
field ${\bf u}({\bf x},t)$, follow the same path as fluid molecules and
evolve according to the dynamics $ \dot{\bf X}(t) = {\bf v}(t) \equiv
{\bf u}({\bf X}(t),t) $, where the Lagrangian velocity ${\bf v}$ equals
the Eulerian one ${\bf u}$ computed at the particle position ${\bf
X}$. Such particles constitute a clear-cut indicator of the underlying
turbulent fluctuations. Recently, it has been shown, by comparing the 
different numerical studies and different experimental
results  \cite{arneodoetal08,bifetal07}, that
Lagrangian turbulence is universal, intermittent, and well described
by a suitable generalization of the Eulerian Multifractal formalism to
the Lagrangian domain \cite{Bo.93,bof02,CRLMPA.03}.

Lagrangian Structure Functions (LSF) are defined as:
\begin{equation}
{\cal S}_{{\rm L},i}^{(p)}(\tau) \equiv \langle [v_i(t+\tau) -v_i(t)]^p \rangle=
\langle (\delta_\tau v_i)^p \rangle, 
\label{eq:LSF}
\end{equation} 
where $i=x,y,z$ runs over the three velocity components, and the
average is defined over the ensemble of particle trajectories evolving
in the flow.   From now on, we will assume isotropy and therefore drop
the dependency from the spatial direction $i$. 
The presence of long spatial and temporal
correlations suggests, in analogy with critical phenomena, 
the existence of scaling laws for time scales larger than the
dissipative Kolmogorov time  and smaller than the typical large-scale
time,  $\tau_\eta \ll \tau \ll T_{\rm L}$:xs
\begin{equation}
{\cal S}_{\rm L}^{(p)}(\tau)  \sim \tau^{z(p)}\,.
\label{eq:scaling}
\end{equation}
Straightforward dimensional arguments \textit{\`a la} Kolmogorov
predict $z(p)=p/2$, independently of the flow properties.  However,
it is known that LSF experience show strong variations when changing the time
lags $\tau$, as highlighted by the increasingly non-Gaussian tails
characterizing the probability density functions of $\delta_\tau v$
for smaller and smaller $\tau$'s \cite{MMMP.01}. This leads to a
breakdown of the dimensional argument; correspondingly, there is a
  growth
of the Lagrangian flatness when
 going to smaller and smaller time lags 
In \cite{arneodoetal08,bifetal07}
it has been shown that the scaling exponent $\kappa(4) = z(4)/z(2)$,
entering in the evolution of the fourth-order flatness:
\be
\frac{{\cal S}_{\rm L}^{(4)}(\tau)}{({\cal S}_{\rm L}^{(2)}(\tau))^2} 
\sim ({\cal S}_{\rm L}^{(2)}(\tau))^{\kappa(4)-2} 
\ee
does not depend on the experimental or numerical large-scale
set-up. In other words, the high frequency fluctuations are universal. 

It is possible to obtain a link between Eulerian and Lagrangian MF formalism 
via the dimensional relation \cite{Bo.93,bof02}: 
\be
\label{eq:lagr}
\tau \sim r/\delta_r u .
\ee
Indeed, if we substitute (\ref{eq:lagr}) into  (\ref{eq:pf}) and into (\ref{zetap})
 we obtain a  Lagrangian prediction for LSF once the Eulerian one
is known via its $F(h)$ spectrum 
\cite{bof02}, namely
\be
\label{eq:lsf}
{\cal S}_{\rm L}^{(p)} (\tau) \sim \int dh \tau^{(ph + F(h))/(1+h)} \sim \tau^{z(p)},
\ee
where 
$$z(p) = \min_h[(ph+F(h))/(1+h)].$$
This result is  obtained  by the saddle
point method for inertial-range time intervals: $ \tau_{\eta} \ll \tau
\ll T_L$. 
Let us notice that the
 above relation (\ref{eq:lsf}) can be read as a prediction for the
 Lagrangian domain, once the Eulerian statistics is known. This is
 because we are using the same MF functions $F(h)$ for both
 domains. The suggested road
 map should be the following: (i) first measure the Eulerian scaling
 exponents $\zeta(p)$; (ii) then via an inverse Legendre transform
 extract the $F(h)$-spectrum; (iii) finally, apply the relation
 (\ref{eq:lsf}) and calculate the Lagrangian scaling. Such procedure
 is working well, at least within the statistical limitation and the
 Reynolds number limitations allowed by  numerical and experimental
 state-of-the-art techniques \cite{arneodoetal08} (see also \cite{zymin} for a
 recent theoretical attempt).

\begin{figure}
\centering
  \includegraphics[width=0.95\textwidth]{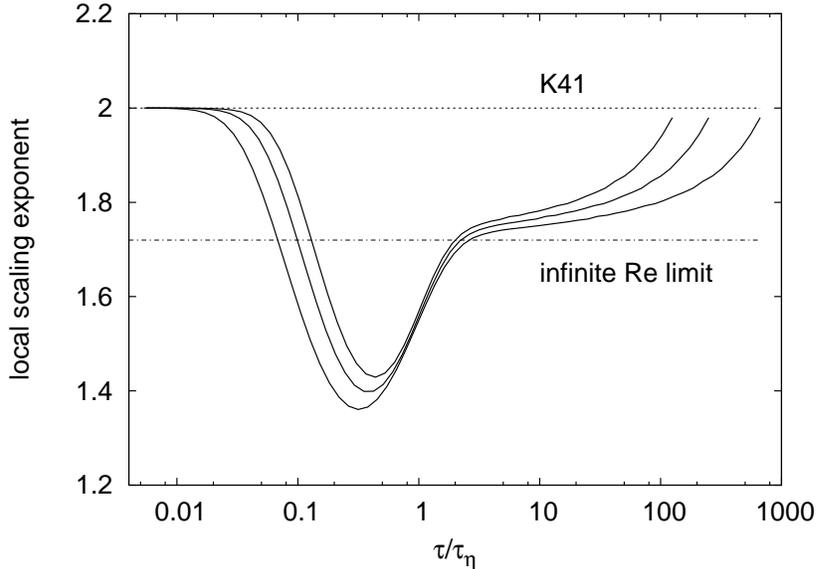}
\caption{Fourth-order local scaling exponents, $\kappa(4,\tau) \equiv
  \frac{d \ \log ({\cal S}_{\rm L}^{(4)}(\tau)}{d  \ \log ({\cal
      S}_{\rm L}^{(2)}(\tau))}$, from the  MF formalism (\ref{eq:lsf}) for
  three different Reynolds numbers, $Re = 2. 10^4, 9.10^4,7.10^5$
  (from top to bottom) The two horizontal line correspond respectively
  to the non intermittent, K41 case, $\kappa(4)=2$,  and to the
  infinite Reynolds number limits, $\kappa(4)=1.72$. The $F(h)$
  spectrum used to derive this values has a  log-Poisson shape as
  proposed  in \cite{sl}. The value of the free parameter is
  $\beta=4$.  Notice that the inertial range extension, identified as
  the region where the local exponent is constant,  becomes larger and
  larger when increasing the Reynolds number. Notice also the strong increase in the intermittency, measured by the deviation from the K41 value, $\kappa=2$, for time scales across the viscous domain. }
\label{fig:1}
\end{figure}

 Even more interesting, in \cite{arneodoetal08}, the same argument, leading to the spatial
 dissipative fluctuating scale (\ref{fv}), has been extended  into the Lagrangian domain
 to obtain an expression for the fluctuating 
 dissipative time scale \cite{BBCDLT.04}:
\begin{equation}
\label{eq:visco}
\tau_{\eta}(h) \sim \nu^{(1-h)/(1+h)} \,.
\end{equation}
Following the same ideas discussed in Sec.~\ref{sec:mf} about the 
viscous modification to the MF formalism, we may now introduce  two functions,
 ${\tilde f}(\frac{\tau}{\tau(h)})$ and ${\tilde
  g}(\frac{\tau}{\tau(h)})$,
which  take into account the viscous corrections to the Lagrangian inertial scale (\ref{eq:lsf}).
Is is customary to use
for the two cross-over functions a Batchelor-Meneveau 
functional form  \cite{meneveau,CRLMPA.03,arneodoetal08}.
 The global description for  
velocity increments becomes then:
\begin{equation}
\label{eq:fit}
\delta_{\tau} v(h,\tau_\eta(h))  \sim  \frac{\tau/T_L}{[(\frac{\tau}{T_L})^{\beta} + 
(\frac{\tau_{\eta}(h)}{T_L})^{\beta}]^{\frac{1-2h}{\beta(1-h)}}}\,,
\end{equation}
in which $\beta$ is a free parameter which controls the crossover from
dissipative to inertial time lags. It is easy to realize that the
above expression reproduces the asymptotic regimes, $\delta_\tau v
\sim \tau $ for $\tau \ll \tau_\eta$ and $\delta_\tau v \sim
\tau^{h/(1-h)}$ in the inertial range.  As in the Eulerian case, each
exponent $h$ must be weighted with its probability, which is also
modeled to include dissipative-range physics, mimicking what has
already been discussed in the introduction for the Eulerian case:
\begin{equation}
P_h(\tau,\tau_{\eta}(h)) = {\cal Z}^{-1}(\tau) \left[ \left(\frac{\tau}{T_L}\right)^{\beta} + 
\left(\frac{\tau_{\eta}(h)}{T_L}\right)^{\beta}\right]^{\frac{F(h)}{\beta(1-h)}}.
\end{equation}
Here ${\cal Z}$ is a normalizing factor. We notice that again 
everything is expressed in terms of the same MF function $F(h)$. Only a new free parameter, $\beta$, enters to describe the whole Lagrangian behavior, at all time scales, once the Eulerian MF spectrum, $F(h)$, is known.  
Putting everything  together, we obtain for the LSF the following MF prediction:
\begin{equation}
\label{eq:lsfmulti}
\langle (\delta_{\tau} v)^p \rangle \sim  \int dh P_h(\tau,
\tau_{\eta}) [\delta_{\tau} v(h)]^p\,.
\end{equation}
The previous formula for the Lagrangian domain predicts a highly non
trivial shape for the local scaling exponents entering in the Flatness
behavior:
\be
\label{eq:lse}
\kappa(4,\tau) \equiv  \frac{d \ \log ({\cal S}_{\rm L}^{(4)}(\tau)}{d  \ \log ({\cal S}_{\rm L}^{(2)}(\tau))}.
\ee
As shown in fig.~(\ref{fig:1}), where we plot it for a series  of
different Reynolds numbers and for a given choice of the multifractal
spectrum $F(h)$ (see caption for details). 
%%%%% FIGURE intro1 %%%%%%%%%%%%%%%
%%%%%%%%%%%%%%%%%%%%%%%%%%%%%%%%%%
The functional dependence shown in fig.~(\ref{fig:1})
was tested in
\cite{arneodoetal08} only for $\kappa(4)$  and for a limited
set of Reynolds numbers, due to the unavoidable limitations in
experiments and numerical simulations. We want now to test it further, by
boosting the Reynolds number by orders of magnitudes. To achieve this, we
switch to a  dynamical surrogate of the Navier--Stokes
equations. 

\subsection{Shell Models}
\label{sec:sm}
Let us now discuss in more details the behavior of the local scaling
exponents for the Lagrangian structure functions (\ref{eq:lse}).  A
decrease of the local scaling exponent simply means that intermittency
is growing in the dissipation range. A detailed analysis of the flow
configurations, which can be performed in high-resolution numerical
simulations, show that the tip in the scaling exponents is strongly
related to the presence of coherent three-dimensional vortices
\cite{BBCLT.05,BBCLT.06}.  This is an important observation which
needs to be investigated carefully. The physical question we are
discussing, concerns the possibility that coherent structures, i.e.
vortices, are responsible for intermittency in three-dimensional
turbulence, not just in the dissipative region but for the whole
inertial range. After all, if the largest intermittent fluctuations of
velocity gradients are due to vortices, it is somehow reasonable to
think of the effect of coherent structure as a key feature for
explaining intermittency. If this is true, the multifractal conjecture
may be misleading, i.e. scale invariance is not a fundamental
properties of three-dimensional turbulence.  Nevertheless, in
\cite{arneodoetal08} it has been shown that the multifractal
conjecture does predict the tip in the local scaling exponents for the
structure functions.

So far, we have discussed the statistical properties of homogeneous
and isotropic turbulence by checking the multifractal predictions
against numerical and experimental data.  In the last twenty years, it
has been shown that there exists a class of simplified models, named
shell model, which shows multifractal intermittency (anomalous
scaling) similar qualitatively and quantitatively to what it is
observed in  Navier--Stokes turbulence.  Among many different shell
models, we shall consider the shell model proposed in  \cite{sabra} ( see also \cite{lb.arfm} for a review).
Shell models of turbulence, can be seen as  a truncated
description of the Navier--Stokes dynamics, preserving some of the
structure and conservation laws of the original  equations but {\it destroying} all spatial structures. They are described by the following set of ODEs:
\begin{eqnarray}
(\frac{d }{dt}&+&\nu k_n^2 ) u_n = i(k_{n+1} u_{n+1}^* u_{n+2}-\delta
k_n u_{n-1}^* u_{n+1}\nonumber\\&+&(1-\delta) k_{n-1} u_{n-1} u_{n-2})+f_n \ . \label{sabra}
\end{eqnarray}
Here the $u_n$s are the velocity modes restricted to `wavevectors' $k_n=k_0
2^n$ with $k_0$ determined by the inverse outer scale of
turbulence. The model contains one free parameter, $\delta$, and
it conserves two quadratic invariants (when the force and the
dissipation term are absent) for all values of $\delta$. The first is
the total energy $\sum_n |u_n|^2$ and the second is  a sort of generalized 
{\it helicity} 
$\sum_n (-1)^n k_n^{\alpha} |u_n|^2$, where $\alpha= \log _2 (1-\delta)$ \cite{lb.arfm}.  Here we consider values of the parameters such
that $0<\delta<1$.
The scaling exponents characterize the shell model
structure functions, defined as
\begin{eqnarray}
S^{(2)}(k_n)\equiv \langle u_n u^*_n\rangle \sim k_n^{-\zeta(2)} \ , \label{S2}\\
S^{(3)}(k_n)\equiv {\rm Im}\, \langle u_{n-1}u_n u_{n+1}^*\rangle \sim k_n^{-\zeta(3)}\ ,\label{S3}\\
S_p(k_n)\sim k_n^{-\zeta(p)} \  . \nonumber
\end{eqnarray}
The values of the scaling exponents have been determined accurately by
direct numerical simulations. Besides $\zeta(3)$ which is exactly
unity, because a relation similar to (\ref{K45}) holds, 
 all the other exponents $\zeta(p)$ are anomalous, differing
from $p/3$.  For $\delta = -0.4$, the value of the $\zeta(p)$ are close to the scaling exponents of the Navier--Stokes equations.
In applying the multifractal conjecture to the shell model, we shall assume $u_n \sim k_n^{-h}$ with probability
$k_n^{-F(h)}$, while the fluctuating dissipative scale $k_d(h)$ is defined by the relation $u_d/(k_d\nu) \sim 1$, where $u_d = k_d^{-h}$.
In the dissipation range, the behavior of the shell model is roughly
consistent with 
 $u_n \sim k_n \exp (-k_n/k_d)$.\footnote{Balancing of the nonlinear and viscous terms
  in the far dissipation range actually gives to leading order $u_n
  \sim k_n \exp(- k_n^\alpha)$ with $\alpha \simeq 0.69$ solution of
$2^{-\alpha} +4^{-\alpha} = 1$.}
By matching between the  inertial range (i.e. $u_n \sim k_n^{-h}$) and
the dissipation range, we then obtain:
\begin{equation}
\label{eq:mf1}
u_n \sim k_n^{-h}(1+ A(k_n/k_d(h))^{1+h})\exp(-k_n/k_d(h))
\end{equation}
where $A$ is a Reynolds independent quantity. Consequently, the probability distribution $P_h(k_n)$ should be modified in order
to take into account the dissipation effects, namely:
\begin{equation}
\label{eq:mf2}
P_h(k_n) \sim [k_n(1+A(k_n/k_d(h)))^{-1}]^{-F(h)}.
\end{equation}
Knowing $F(h)$, which as usual can be estimated from the knowledge of $\zeta(p)$, we can compute the $k_n$ dependency of $S^{(p)}(k_n)$ both
in the inertial and in the dissipative range and compare our findings with numerical simulations of (\ref{sabra}).

\begin{figure}
\centering
  \includegraphics[width=0.95\textwidth]{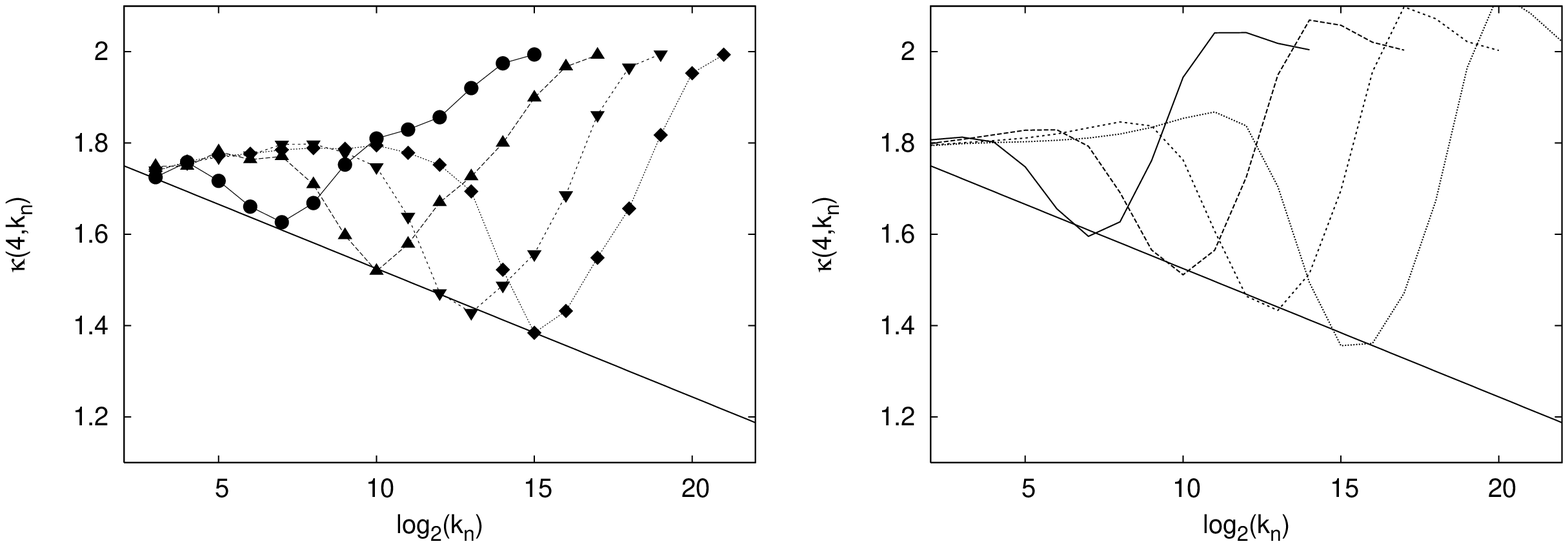}
  \includegraphics[width=0.95\textwidth]{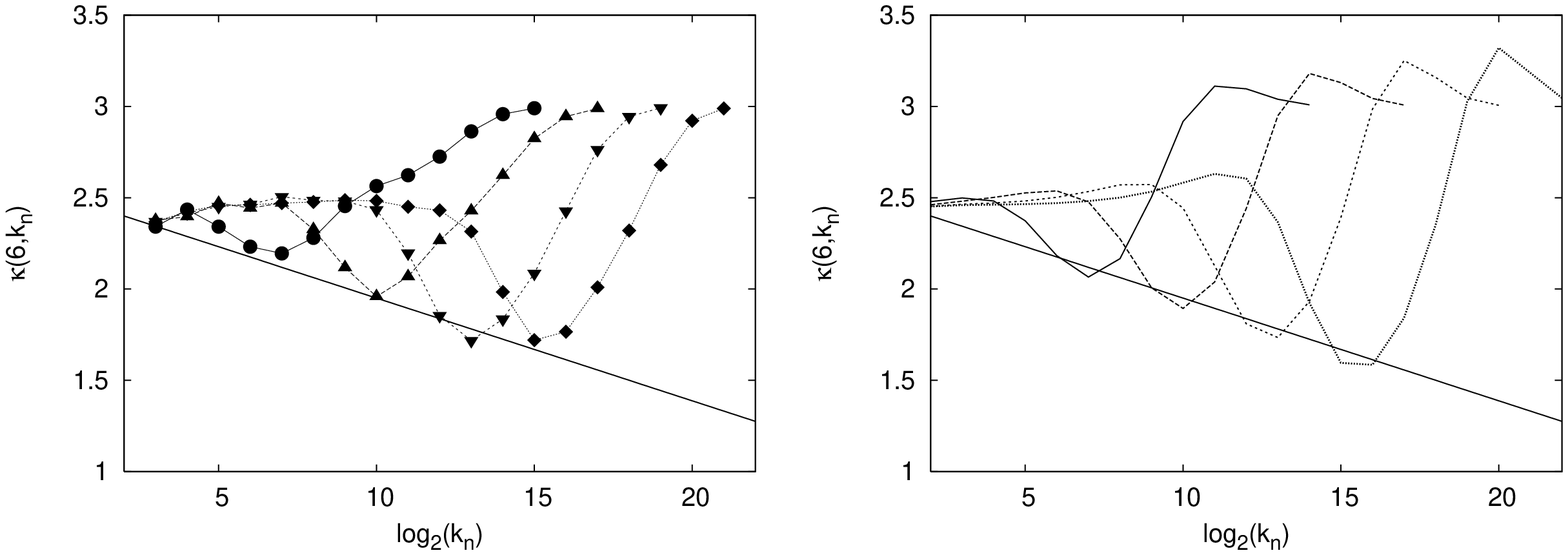}
\caption{Top: local scaling exponents for fourth order structure functions,
$\kappa(4,k_n)$, in the Shell Model, versus $\log _2(k_n)$. Left numerical results at  four Reynolds numbers in the range $Re \in [10^2:10^4]$. Right, the MF prediction using relations (\ref{eq:mf1}-\ref{eq:mf2}); the straight line is a fit for the behavior of the bottleneck (maximum  of intermittency), the slope of the line is  $-0.028$.
 Bottom: the same for the sixth order scaling exponents $\kappa(6, k_n)$. The straight line has the slope $-0.056$}
\label{fig02}
\end{figure}

In fig.~(\ref{fig02}) we plot the local scaling exponents 
$$
\kappa(4,k_n) \equiv  \frac{d \ \log (S_4(k_n)}{d  \ \log (S_2(k_n))}.
$$
computed from the numerical simulations of the shell model (upper
left panel) and predicted by the multifractal conjecture (upper right
panel). Different symbols refer to different Reynolds numbers. The
first striking result is that the numerical simulations of the model
clearly show a well defined tip in the dissipative region (i.e. large
value of $n$ in the figure), similar to what it is observed in
experiments and in the numerical simulations. The tip is increasing
towards small scales as the Reynolds number increases and it
deepens. The straight line in the figure is a qualitative fit on the
behavior of the tip as a function of $Re$. Clearly, the increase of
intermittency in the dissipative region is scaling as
$\log (Re)$. Notice that this scaling behaviour was not visible in the
experimental and numerical data shown in \cite{arneodoetal08,bifetal07}
because of the limited range of Reynolds spanned in those cases.  In
the right  panel, we show $\kappa(4,k_n)$ as predicted by using the
multifractal conjecture with $A=0.8$. As one can observe, the
qualitative and quantitative predictions are in remarkable agreement
with the numerical simulations.

%%%%% FIGURE intro1 %%%%%%%%%%%%%%%
%%%%%%%%%%%%%%%%%%%%%%%%%%%%%%%%%%

The quality of the result does not change by increasing the order of
the structure functions. In fig.~(\ref{fig02}) we also show behavior
of $\kappa(6,k_n)$ (i.e. the local scaling exponents of the sixth order
structure functions) and compare the numerical simulations with the
multifractal prediction. Again, the results confirms what we found for
the lower order. Notice again, the widening of the inertial range when
increasing the Reynolds number  and the power-law behavior of the
wavenumber where we observe the highest intermittency, i.e. the
minimum in the local scaling exponents.

From the above analysis we can draw some interesting
conclusions. First of all, the increase of intermittency in the
dissipative range is not due to coherent vortices (there are no
vortices in the shell model). Moreover, the increase of intermittency
is predicted by the multifractal conjecture because of the
fluctuations of the dissipative scale. Translating back these results
to the real Navier--Stokes equation, we are tempted to conclude that
coherent structures exist but their dynamics is nor relevant to
explain intermittency in turbulent flows.  They are the tail more than
the dog (see chapter 8.9 of \cite{Fr.95}). Actually, one can take the
opposite point of view: the matching between inertial range (scale
invariance) intermittency and the dissipation range produces an
increase in fluctuations and, consequently, an increase of
vorticity. This effect is dominant at low Reynolds number, as it is
clearly observed in the shell model: i.e. at low Reynolds numbers no
scaling behavior is observed and intermittency is strongly dominated
by the fluctuations in the dissipation range. Let us also point out
that our observations is in qualitative agreement to the known
phenomenology of boundary layer turbulence. Near the wall, the local
Reynolds number is relatively low and strong intermittency is observed
together with a rich dynamic behavior of coherent structures (hairpin
vortices). The full dynamics is dominated by strong intermittent
fluctuations and it is tempting to relate our previous discussion to
this specific well known turbulent flows.

What is
remarkable in the above discussion,  is that the change in the intermittency level, is explained by the
same MF spectrum $F(h)$, at all scales and at all Reynolds numbers. In
other words, the study of low Reynolds turbulence, and of the transition
from viscous to inertial range \cite{lohse,lohse1}, may teach us a lot
with respect to the high Reynolds limits \cite{yakhot.sreene}.

\section{Conclusions and Perspectives}
Let us summarize our main points. First, we have reviewed the ideas
about how to introduce, in a self-consistent way, dissipative effects in
the Multifractal description of turbulence, both Eulerian and
Lagrangian. The MF formalism, predicts an enhancement of intermittency
in the so-called intermediate viscous range \cite{FV.91}, as measured
by the local scaling exponents (see fig,~\ref{fig:1}). We have
commented on the fact that such a trend is in very good agreement
with real Lagrangian Turbulence data \cite{arneodoetal08}, at least
concerning low order moments and moderate Reynolds numbers, i.e. up to
the numerical and experimental state-of-the-art.  Second, in order to
test the formalism also for high Reynolds numbers, we switched to a
class of Shell Models of turbulence. Here, thanks to the much simpler
structures of the model, the Reynolds dependence of the MF prediction
can be also tested. We have specialized  our discussion  on the enhancement of intermittency measured around
dissipative scales.  The existence of the very same phenomenon
observed in real Navier--Stokes eqs, lead us to conclude that the
viscous intermittency is not due to vortex filaments. Many problems
remain opened. For example, the Batchelor-Meneveau structure presented
in (\ref{eq:fit})  is not compatible with the requirement that
velocity fluctuations follow a simple multiplicative local --in
scales-- process from large to small scales. This is due to the fact
that the fluctuating temporal  scale $\tau_\eta(h)$
appears in the definition of the velocity increments and it depends on
them (in eqs. (\ref{eq:visco}-\ref{eq:fit}) the local scaling
exponent $h$ is the same). In other words, the functional relation
(\ref{eq:fit}) introduces non-local correlation between inertial and
dissipative scales. A local-in-scale multiplicative process which
takes into account the fluctuating cutoff can be introduced somewhat
empirically by building a multiplicative cascade and stopping it
according to the criteria ({\ref{fv}).  It would be interesting to see
if such a procedure reproduces the Navier--Stokes data and the Shell
Model data as nicely  as  (\ref{eq:fit}).

We acknowledge many  fruitful collaborations on these issues with
M. Cencini, A.S. Lanotte and F. Toschi. L.B. wants to acknowledge also fruitful discussions with V. Yakhot.

\end{document}